%% file: ieee_conf.tex
\documentclass[conference]{IEEEtran}
\IEEEoverridecommandlockouts
\usepackage{cite}
\usepackage{amsmath,amssymb,amsfonts}
\usepackage{algorithmic}
\usepackage{graphicx}
\usepackage{textcomp}
\usepackage{booktabs}
\usepackage{xcolor}
\def\BibTeX{{\rm B\kern-.05em{\sc i\kern-.025em b}\kern-.08em
    T\kern-.1667em\lower.7ex\hbox{E}\kern-.125emX}}

\usepackage{caption} %
\usepackage{subcaption} %
\usepackage{bm}

\usepackage{newfloat}    
\DeclareFloatingEnvironment[
fileext=lop,
name=Model,
placement=p,
]{model}

\begin{document}

\title{
Second-Order-Cone Formulations of Power Flow for Topology Optimization\\
\thanks{N.~Rhodes and L.A.~Roald acknowledge support from the National Science Foundation (NSF) under the NSF CAREER award No. 2045860 and NSF ASCENT award No. 2132904. This work is supported in part by the U.S. Department of Energy
through the LANL/LDRD Program and the Center for Nonlinear Studies. 
This research used resources provided by the Los Alamos National Laboratory Institutional Computing Program, which is supported by the U.S. Department of Energy National Nuclear Security Administration under Contract No. 89233218CNA000001.}
}

\author{\IEEEauthorblockN{Noah Rhodes}
\IEEEauthorblockA{\textit{Los Alamos National Laboratory} \\
Los Alamos, New Mexico, USA \\
nrhodes@lanl.gov}
\and
\IEEEauthorblockN{James Luedtke}
\IEEEauthorblockA{\textit{Industrial and Systems Engineering } \\
\textit{University of Wisconsin - Madison}\\
Madison, Wisconsin, USA \\
jim.luedtke@wisc.edu}
\and
\IEEEauthorblockN{Line Roald}
\IEEEauthorblockA{\textit{Electrical and Computer Engineering} \\
\textit{University of Wisconsin - Madison}\\
Madison, Wisconsin, USA \\
roald@wisc.edu}
}

\maketitle

\begin{abstract}
Optimization problems that involve topology optimization in scenarios with large scale outages, such as post-disaster restoration or public safety power shutoff planning, are very challenging to solve. 
Using simple power flow representations such as DC power flow or network flow models 
results in low quality solutions which requires significantly higher-than-predicted load shed to become AC feasible. Recent work has shown that formulations based on the Second Order Cone (SOC) power flow formulation find very high quality solutions with low load shed, but the computational burden of these formulations remains a significant challenge.  
With the aim of reducing computational time while maintaining high solution quality, this work explores formulations which replace the conic constraints with a small number of linear cuts. 
The goal of this approach is not to find an exact power flow solution, but rather to identify good binary decisions, where the power flow can be resolved after the binary variables are fixed. 
We find that a simple reformulation of the Second Order Cone Optimal Power Shutoff problem can greatly improve the solution speed, but that a full linearization of the SOC voltage cone equation 
results in an overestimation of the amount of power that can be delivered to loads. 
\end{abstract}

\section{Introduction} \label{sec:intro}
Topology optimization for power grids with large scale outages is an active research area that supports important applications such as restoration planning \cite{castillo2014risk, qiu2017integrated} or planning of Public Safety Power Shutoffs (PSPS) for the reduction of wildfire risk \cite{rhodes2020balancing}.  In this work, we focus on on PSPS as an application, but our results are applicable to any topology optimization problem where a significant number of lines are outaged. In those settings, prior work has identified that DC power flow problem formulations produce solutions which have very poor AC-feasible load delivery \cite{mld, van2015transmission, haag2024long,rhodes2021powermodelsrestoration}.

PSPS planning has been widely studied in recent years as the Optimal Power Shutoff (OPS) optimization problem \cite{rhodes2020balancing}.  There are many variations of the problem to account for fairness in load shed \cite{kody2022sharing}, security constraints \cite{rhodes2024security}, or incorporating restoration planning \cite{rhodes2023co}. Many works include investment planning for hardening power-lines \cite{taylor2022framework, bertoletti2022transmission, bayani2023resilient}, building energy storage \cite{kody2022optimizing, astudillo2022managing}, and distributed energy resources~\cite{yang2022resilient, hanna2021optimal}.  Many extensions also include stochastic optimization to account for the uncertainty in wildfire ignition and spread \cite{yang2024multistage,yang2024multi,zhou2024mitigating}. The AC power flow representation is used for PSPS planning \cite{hong2022data}, but it only considers up to three line de-energizations.   All other works mentioned above use a DC power flow formulation.

In \cite{haag2024long}, authors show that using a DC power flow formulation can have very poor performance for the OPS problem, where the decisions made with a DC power flow formulation result in much less power delivered after an AC-feasible power flow is recovered. Solving the OPS problem with the Second Order Cone (SOC) power flow formulation results in very high quality solutions, but it comes at the cost of increased computational runtime. 

In this work, we attempt to find high-quality solutions with reduced runtime by developing a linear relaxation of the SOC formulation of the OPS problem.  In particular, we are interested in finding good binary decisions for the systems topology, and subsequently solving an AC power flow continuous problem to find the load delivery of the system. The relaxation does not have to be accurate, as long as it still leads to good binary decisions.

Our primary contributions are
(1) we improve the SOC performance by reformulating the voltage cone equations, 
(2) we develop two fully linear relaxations of the SOC equations, and 
(3) we perform extensive computational testing to demonstrate the quality and performance trade-offs of the formulations.  

The remainder of the paper is organized as follows: Section \ref{sec:model} Introduces the SOC OPS modeling, and \ref{sec:linearize_quads} develops the linearized formulations.
Section \ref{sec:case_study} describes the case study setup.  The results are presented in Section \ref{sec:results} with a discussion in Section \ref{sec:discussion}. The work is concluded in Section \ref{sec:conclusion}.

\section{Topology Optimization with Exact Second-Order Cone Constraints} \label{sec:model}
In this section, we first describe the SOC-OPS model as implemented by \cite{haag2024long}, which uses the topology switching SOC model from~\cite{mld}.  We then present an exact reformulation of the problem that improves the voltage-cone equations and the bounds on the shunt voltage.  

The sets of lines, buses, generators, load demand and shunts are represented by $\mathcal{L}$, $\mathcal{B}$, $\mathcal{G}$, $\mathcal{D}$, and $\mathcal{S}$, respectively.  Sets of components at a bus, such as the generators at bus $i$ are represented as $\mathcal{B}_i^{\mathcal{G}}$.  Parameters are in bold script, while variables are in normal script.  The component energization variables, $z$, are binary and all other variables are continuous.

\subsection{SOC OPS Formulation} \label{sec:SOC_OPS}
We first introduce the SOC-OPS model from \cite{haag2024long}.  The objective is specific to the  model in \cite{haag2024long}, whereas the rest of the formulation is applicable to any power flow and topology optimization problem.

\noindent \emph{Objective Function:}
The objective function for the OPS problem, shown in eq. \eqref{eq:obj}, maximizes load delivery while minimizing the wildfire risk on energized power lines, with the competing objectives balanced by the parameter $\boldsymbol{\alpha} \in [0,1]$.  The term $\boldsymbol{P}_d^D$ is the power demand of $d$, which is multiplied by the continuous variable $x_d \in [0,1]$, representing the percentage of the demand that is delivered. The load is multiplied by a weight parameter $\boldsymbol{w}_d$ which allows critical loads to be prioritized.  The load delivery is summed over all loads, and divided by the total load demand of the system $\boldsymbol{P}_{tot}^D$ to normalize that term of the objective.

The second objective term represents the total risk on energized power lines.  $\boldsymbol{R}_{ij}$ is the wildfire risk of line $ij$, and $z_{ij}$ is a binary variable representing the energization state of the power line.  A line is only able to ignite a wildfire if it is energized, and the risk is $0$ when the line is de-energized, i.e., when $z_{ij}=0$.  The risk is summed over all energized lines, and divided by the total risk of all power lines $\boldsymbol{R}_{tot}$ to normalize the risk component of the objective: 
\begin{equation}
    \max \;\; (1-\boldsymbol{{\alpha}})\frac{\sum_{d\in\mathcal{D}} x_d \boldsymbol{w}_d \boldsymbol{P}^D_d}{\boldsymbol{P}^D_{tot}} - \boldsymbol{{\alpha}}\frac{\sum_{ij\in\mathcal{L}}z_{ij} \boldsymbol{R}_{ij}}{\boldsymbol{R}_{tot}} . \label{eq:obj}
\end{equation}

\noindent \emph{Energization Constraints:}
The following constraints represent the energization state of the components, and limit how energized components may by connected to de-energized components.  As an example, in eq. \eqref{eq:ops_gen_active} generator $g$ must be de-energized if the bus it connects to $i$ is de-energized, but the bus may be energized while the generator  is de-energized.  The same constraints apply for all other components that connect to a bus, as shown in eqs. \eqref{eq:ops_relationships}. The energization state of a branch has two constraints because it has connections to two buses: 
\begin{subequations}
\begin{align}
        &  z_{g} \le z_{i}  &&  \forall g \in \mathcal{B}^\mathcal{G}_i, \; \forall i \in \mathcal{B}   \label{eq:ops_gen_active}\\
        &  z_{i j } \le z_{i} &&  \forall ij \in \mathcal{B}_i^{\mathcal{L}}, \; \forall i \in \mathcal{B}   \label{eq:ops_line_active_i}\\
        &  z_{i j } \le z_{j} &&  \forall ij \in \mathcal{B}_i^{\mathcal{L}}, \; \forall i \in \mathcal{B}   \label{eq:ops_line_active_j}\\
        &  x_{d} \le z_{i}  &&   \forall d \in \mathcal{B}^\mathcal{D}_i, \; \forall i \in \mathcal{B}  \label{eq:ops_load_active} \\
        &  x_{s} \le z_{i}  &&   \forall s \in \mathcal{B}^\mathcal{S}_i, \; \forall i \in \mathcal{B} .\label{eq:ops_shunt_active}
\end{align}
\label{eq:ops_relationships} 
\end{subequations}
The bounds of the energization state variables are shown in eq. \eqref{eq:z_bounds}.  The state of the generators, lines, and buses are binary while the state of the loads and shunts can be continuously shed:  
\begin{subequations}
    \begin{align}
        & z_g \in \{0,1\} & \forall g \in \mathcal{G} \label{eq:zg_bound} \\
        & z_{ij} \in \{0,1\} & \forall ij \in \mathcal{L} \label{eq:zij_bound} \\
        & z_i \in \{0,1\} & \forall i \in \mathcal{B} \label{eq:zi_bound} \\
        & 0 \le x_d \le 1 & \forall d \in \mathcal{D} \label{eq:xd_bound} \\
        & 0 \le x_s \le 1 & \forall s \in \mathcal{S} . \label{eq:xs_bound} 
    \end{align}
    \label{eq:z_bounds}
\end{subequations}
\noindent \emph{Generation Constraints:}
The generation limits are shown in \eqref{eq:ops_active_gen_limits} and \eqref{eq:ops_reactive_gen_limits}.  When a generator is energized, the generator output for real power $P_g^G$ and reactive power $Q_g^G$ is limited to the maximum and minimum power and reactive power limits, $\underline{\boldsymbol{P}^G_g}$, $\overline{\boldsymbol{P}^G_g}$, $\underline{\boldsymbol{Q}^G_g}$, and  $\overline{\boldsymbol{Q}^G_g}$ respectively. The generators cannot output real or reactive power when de-energized, and this is enforced by multiplying the power limits by $z_g$:
\begin{equation}
        z_{g}\underline{\boldsymbol{P}^G_g} \le P_{g}^G \le z_{g} \overline{\boldsymbol{P}^G_g} \quad \forall g \in \mathcal{G}, \label{eq:ops_active_gen_limits}
\end{equation}
\begin{equation}
        z_{g}\underline{\boldsymbol{Q}^G_g} \le Q_{g}^G \le z_{g} \overline{\boldsymbol{Q}^G_g} \quad \forall g \in \mathcal{G} .\label{eq:ops_reactive_gen_limits}
\end{equation}
\noindent \emph{Power Flow Constraints:}
The real power balance at each bus is enforced at each bus by eq. \eqref{eq:ops_soc_active_power_balance}.  The total power from generators $P_g^G$, power lines flows $P_{ij}^L$, load $x_d \boldsymbol{P}^D_d$, and shunts must sum to $0$.  The shunt real power is determined by multiplying the bus shunt conductance $\boldsymbol{g}_i$ by the shunt squared voltage variable $W^S_{s}$.   The shunt voltage variable $W^S_{s}$ represents the term $x_s W_{ii}$, where $W_{ii}$ is the squared voltage at the bus, and this relation is relaxed via the linear McCormick inequalities \eqref{eq:ops_soc_shunt_relaxation}.     
The reactive power balance is similarly modeled in eq. \eqref{eq:ops_soc_reactive_power_balance} using the reactive power variables for generators, lines, and load, with the shunt susceptance $\boldsymbol{b}_i$ for reactive power in shunts:  
\begin{equation}
        \sum_{g\in\mathcal{B}_i^\mathcal{G}}P_{g}^G - \!\!\!\!\!\! \sum_{(i, j)\in\mathcal{B}_i^\mathcal{L}} \!\!\!\!\!  P_{i j}^L - \!\!\!\!  \sum_{d\in\mathcal{B}_i^\mathcal{D}} \!\!\! x_{d} \boldsymbol{P}^D_d  - \boldsymbol{g}_i W^S_s = 0 \quad \forall i \in \mathcal{B},  \label{eq:ops_soc_active_power_balance} 
\end{equation}
\begin{equation}
        \sum_{g\in\mathcal{B}_i^\mathcal{G}}Q_{g}^G - \!\!\!\!\!\! \sum_{(i, j)\in\mathcal{B}_i^\mathcal{L}} \!\!\!\!\!  Q_{i j}^L - \!\!\!\!  \sum_{d\in\mathcal{B}_i^\mathcal{D}} \!\!\! x_{d} \boldsymbol{Q}^D_d + \boldsymbol{b}_i W^S_s = 0 \quad \forall i \in \mathcal{B}. \label{eq:ops_soc_reactive_power_balance} 
\end{equation}
The power flow thermal limit is enforced by eq. \eqref{eq:ops_complex_thermal_limit}, with a constraint for each direction of flow.    The flow is constrained to the thermal limit $\boldsymbol{T}_{ij}$ when the line is energized, and to $0$ when the line is de-energized: 
\begin{subequations}
\begin{align}
        && (P_{i j}^L)^2 + (Q_{i j}^L)^2 \le \boldsymbol{T}_{i j}^2z_{i j} \quad \forall ij \in \mathcal{L} \label{eq:ops_complex_thermal_limit_to} \\
        && (P_{j i}^L)^2 + (Q_{j i}^L)^2 \le \boldsymbol{T}_{i j}^2z_{i j} \quad \forall ij \in \mathcal{L}. \label{eq:ops_complex_thermal_limit_fr}
\end{align} \label{eq:ops_complex_thermal_limit}
\end{subequations}
Power flow on each line is constrained by eq. \eqref{eq:ops_soc_powerflow}, representing real and reactive power flow in each direction on a line.  The power flow is a function of several voltage variables:  $W^{Fr}_{ij}$ is the squared voltage at bus $i$, $W^{To}_{ij}$ is the squared voltage at bus $j$, $W^R_{ij}$ represents the real part of the voltage product of $V_i V_j \cos(\theta_i - \theta_j)$ and $W^I_{ij}$ represents the imaginary part of the voltage product $V_i V_j \sin(\theta_i-\theta_j)$.  The bounds of these variables are defined later in this section, but all have a bound that multiplies by $z_{ij}$. As a result, $z_{ij}$ does not appear in the power flow equations below, because the voltage variables are all equal to $0$ when $z_{ij} = 0$.  In the power flow equations,  the parameters are defined as follows:  $\boldsymbol{g}_i$ is the shunt conductance of the line at bus $i$, $\boldsymbol{b}_i$ is the shunt susceptance of the line at bus $i$, $\boldsymbol{g}_{ij}$ is the conductance of the line, $\boldsymbol{b}_{ij}$ is the susceptance of the line, $\boldsymbol{t}_{ij}$ is the complex transformer tap ratio, $\boldsymbol{t}^R_{ij}$ is the real part of the transformer tap ratio, $\boldsymbol{t}^I_{ij}$ is the imaginary part of the transformer tap ratio.  
\begin{subequations}
\begin{align}
    \begin{split}
        P_{ij}^L &=   
            \frac{\boldsymbol{g}_{ij} + \boldsymbol{g}_{i}}{|\boldsymbol{t}_{ij}|^2} W^{Fr}_{ij} 
            + \! \frac{-\boldsymbol{g}_{ij} \boldsymbol{t}^R_{ij} \!+\! \boldsymbol{b}_{ij} \boldsymbol{t}^I_{ij}}{|\boldsymbol{t}_{ij}|^2} W^R_{ij} \\ 
            & 
            + \! \frac{-\boldsymbol{b}_{ij} \boldsymbol{t}^R_{ij} \!-\! \boldsymbol{g}_{ij} \boldsymbol{t}^I_{ij}}{|\boldsymbol{t}_{ij}|^2} W^I_{ij}
        \quad \forall ij \in \mathcal{L} \label{eq:soc_p_power_flow_fr}
        \end{split}\\
    \begin{split}
        P_{ji}^L &=    
            \left( \boldsymbol{g}_{ij} + \boldsymbol{g}_{j} \right) W^{To}_{ij} 
            + \! \frac{-\boldsymbol{g}_{ij} \boldsymbol{t}^R_{ij} \!-\! \boldsymbol{b}_{ij} \boldsymbol{t}^I_{ij}}{|\boldsymbol{t}_{ij}|^2} W^R_{ij} \\
            & 
            + \! \frac{-\boldsymbol{b}_{ij} \boldsymbol{t}^R_{ij} \!+\! \boldsymbol{g}_{ij} \boldsymbol{t}^I_{ij}}{|\boldsymbol{t}_{ij}|^2} W^I_{ij}
        \quad \forall ij \in \mathcal{L}  \label{eq:soc_p_power_flow_to}
        \end{split}\\
    \begin{split}
        Q_{ij}^L &=    
             -\frac{\boldsymbol{b}_{ij} + \boldsymbol{b}_{i}}{|\boldsymbol{t}_{ij}|^2} W^{Fr}_{ij} 
            - \! \frac{-\boldsymbol{b}_{ij} \boldsymbol{t}^R_{ij} \!-\! \boldsymbol{g}_{ij} \boldsymbol{t}^I_{ij}}{|\boldsymbol{t}_{ij}|^2} W^R_{ij} \\ 
            & 
            + \! \frac{-\boldsymbol{g}_{ij} \boldsymbol{t}^R_{ij} \!+\! \boldsymbol{b}_{ij} \boldsymbol{t}^I_{ij}}{|\boldsymbol{t}_{ij}|^2} W^I_{ij}
         \quad \forall ij \in \mathcal{L} \label{eq:soc_q_power_flow_fr} 
    \end{split}\\
    \begin{split}
        Q_{ji}^L &=    
             - \left( \boldsymbol{b}_{ij} + \boldsymbol{b}_{j} \right) W^{To}_{ij} 
            - \! \frac{-\boldsymbol{b}_{ij} \boldsymbol{t}^R_{ij} \!+\! \boldsymbol{g}_{ij} \boldsymbol{t}^I_{ij}}{|\boldsymbol{t}_{ij}|^2} W^R_{ij} \\ 
            & 
            + \! \frac{-\boldsymbol{g}_{ij} \boldsymbol{t}^R_{ij} \!-\! \boldsymbol{b}_{ij} \boldsymbol{t}^I_{ij}}{|\boldsymbol{t}_{ij}|^2} W^I_{ij}
         \quad \forall ij \in \mathcal{L}  \label{eq:soc_q_power_flow_to}
    \end{split}
\end{align}
\label{eq:ops_soc_powerflow}
\end{subequations}

\noindent \emph{Voltage Constraints}
The voltage constraints represent a significant component of the SOC relaxation of the AC power flow equations.  Because we also have switching constraints on the power line, some additional variables are also introduced to the formulation.  

First, we define the squared voltage variable for each bus in eq. \eqref{eq:ops_soc_voltage_bus}, with bounds that reduce voltage to zero when de-energized: 
\begin{equation}
        z_{i} \boldsymbol{\underline{V}}^2_i \le  W_{ii}  \le  z_{i} \boldsymbol{\overline{V}}^2_i \quad  \forall i \in \mathcal{B} . \label{eq:ops_soc_voltage_bus}
\end{equation}

We introduce two additional variables in eq. \eqref{eq:ops_soc_voltage_fr_to} for each line to represent the nodal squared voltage at each end of the line.  These variables have the same bounds as the nodal voltage variable:
\begin{subequations}
    \begin{align}
        & z_{ij} \boldsymbol{\underline{V}}^2_i \le  W^{Fr}_{ij}  \le  z_{ij} \boldsymbol{\overline{V}}^2_i & \forall ij \in \mathcal{L} \\
        & z_{ij} \boldsymbol{\underline{V}}^2_j \le  W^{To}_{ij}  \le  z_{ij} \boldsymbol{\overline{V}}^2_j & \forall ij \in \mathcal{L}. 
    \end{align} \label{eq:ops_soc_voltage_fr_to}
\end{subequations}

In eq. \eqref{eq:ops_soc_voltage_fr_to_ii_jj}, the new variables $W^{Fr}_{ij}$ and $W^{To}_{ij}$ are constrained to the same value as the respective nodal voltage when the line is energized, but are set to $0$ when the line is de-energized in combination with eq. \eqref{eq:ops_soc_voltage_fr_to}.  This allows the the power flow equations \eqref{eq:ops_soc_powerflow} to avoid multiplying by $z_{ij}$ when using the nodal voltage:
\begin{subequations}
    \begin{align}
        & W_{ii} \ge W^{Fr}_{ij}  \ge W_{ii} - \boldsymbol{\overline{V}}_i^2 (1-z_{ij}) & \forall ij \in \mathcal{L} \\
        & W_{jj} \ge W^{To}_{ij}  \ge W_{jj} - \boldsymbol{\overline{V}}_j^2 (1-z_{ij}) & \forall ij \in \mathcal{L}. 
    \end{align}  \label{eq:ops_soc_voltage_fr_to_ii_jj}
\end{subequations}

Variables $W^R_{ij}$ and $W^I_{ij}$ represent the real and imaginary portion of the complex voltage product $V_i V_j^*$ in the SOC formulation.  
The bounds multiply the energization state of the line to constrain the value to $0$ when the line is de-energized: 
\begin{subequations}
    \begin{align}
& z_{ij}\boldsymbol{\underline{W}}^R_{ij} \le  W^R_{ij}  \le  z_{ij}\boldsymbol{\overline{W}}^R_{ij} \quad \forall ij \in \mathcal{L}  \\
& z_{ij}\boldsymbol{\overline{W}}^I_{ij} \le  W^I_{ij}  \le  z_{ij}\boldsymbol{\underline{W}}^I_{ij} \quad   \forall ij \in \mathcal{L}.
        \end{align} \label{eq:ops_soc_ij_ri_bounds}
\end{subequations}

The values of $W^R_{ij}$ and $W^I_{ij}$ are inherently related to each other by the voltage angle limits, which are imposed by eq. \eqref{eq:ops_soc_angle_limits}:
\begin{equation}
        \tan \left(\boldsymbol{\underline{\theta}}_{ij}\right) W^R_{ij} \le W^I_{ij} \le \tan\left(\boldsymbol{\overline{\theta}}_{ij} \right) W^R_{ij} \quad \forall ij \in \mathcal{L}.
    \label{eq:ops_soc_angle_limits}
\end{equation}

 The traditional SOC voltage  constraint must be multiplied by $z_{ij}$ to represent a de-energized line, resulting in $(W^R_{ij})^2 + (W^I_{ij})^2 \le W_{ii}W_{jj}z_{ij}$ which is an order-three term.  We use a product relaxation of the right-hand side to create eq. \eqref{eq:ops_soc_ii_jj_ri} which is a set of rotated second order cone constraints: 
\begin{subequations}
    \begin{align}
        & \left(W^R_{ij}\right)^2 +  \left(W^I_{ij}\right)^2  \le W_{ii} W_{jj}  & \forall ij \in \mathcal{L} \\
        & \left(W^R_{ij}\right)^2 +  \left(W^I_{ij}\right)^2  \le W_{ii} \overline{\boldsymbol{W_{jj}}} z_{ij} & \forall ij \in \mathcal{L} \\
        & \left(W^R_{ij}\right)^2 +  \left(W^I_{ij}\right)^2  \le \overline{\boldsymbol{W_{ii}}} W_{jj} z_{ij}& \forall ij \in \mathcal{L}. 
    \end{align} \label{eq:ops_soc_ii_jj_ri}
\end{subequations}

Finally, eq. \eqref{eq:ops_soc_shunt_relaxation} is the McCormick relaxation of the $W^S_s = x_s W_{ii}$.  This relaxation allows the power balance constraint to be linear:  
\begin{subequations}
    \begin{align}
        &  W^S_{s} \ge 0  \quad \forall s \in \mathcal{B}^\mathcal{S}_i, & \forall i \in \mathcal{B} \\
        &  W^S_{s} \ge \boldsymbol{\overline{V}}^2_i (x_s -1 ) + W_{ii}  & \forall s \in \mathcal{B}^\mathcal{S}_i, \; \forall i \in \mathcal{B} \\
        &  W^S_{s} \le W_{ii}  \quad \forall s \in \mathcal{B}^\mathcal{S}_i, & \forall i \in \mathcal{B} \\
        &  W^S_{s} \le \boldsymbol{\overline{V}}^2_i x_s  \quad \forall s \in \mathcal{B}^\mathcal{S}_i, & \forall i \in \mathcal{B}.
    \end{align} \label{eq:ops_soc_shunt_relaxation}
\end{subequations}

The complete SOC OPS optimization problem is:
\begin{align}
    &\max &&  \mbox{Objective \eqref{eq:obj}} \nonumber\\
&\mbox{s.t.: \,\,\,}  \tag{SOC-OPS}
&& \mbox{Component relationships:}~\eqref{eq:ops_relationships}, \eqref{eq:z_bounds} \nonumber \\[-2pt]
&&& \mbox{Generation constraints:}~\eqref{eq:ops_active_gen_limits}, \eqref{eq:ops_reactive_gen_limits}  \nonumber\\[-2pt]
&&& \mbox{SOC power flow:}~\eqref{eq:ops_soc_active_power_balance}-\eqref{eq:ops_soc_shunt_relaxation} .
\nonumber
\end{align}

\subsection{Voltage Cone Perspective Formulation} \label{sec:tighten}
Our first significant observation is that the SOC voltage cone in eq. \eqref{eq:ops_soc_ii_jj_ri} can be simplified and strengthened. 
The right hand side of the constraint is $W_{ii}W_{jj}z_{ij}$ and this is relaxed to three cones using the upper bounds of each of the three variables.  Instead, notice that we can do the following.

Multiplying the equation by $z_{ij}$ does not change the value:
$$ W_{ii}W_{jj}z_{ij} = W_{ii}W_{jj}z_{ij}z_{ij}.$$
Remember that the definition of $W^{Fr}_{ij}$ and $W^{To}_{ij}$ is:
$$ W_{ii}z_{ij}= W^{Fr}_{ij}, W_{jj}z_{ij}=W^{To}_{ij}.$$
Thus,
$$W_{ii}W_{jj}z_{ij}z_{ij}= W^{To}_{ij}W^{Fr}_{ij}.$$
Hence, we can reformulate the voltage cone equations \eqref{eq:ops_soc_ii_jj_ri} as:
 \begin{equation}
     \left(W^R_{ij}\right)^2 +  \left(W^I_{ij}\right)^2  \le W^{Fr}_{ij} W^{To}_{ij}  \quad \forall ij \in \mathcal{L} \label{eq:ops_soc_cone_perp}
 \end{equation}
which is a second-order cone constraint.
This reformulation does not require the product relaxation, and hence both reduces the formulation size and improves the quality of the continuous relaxation, because when $z_{ij}$ is relaxed to be continuous, the term 
$W^{Fr}_{ij} W^{To}_{ij}$ in the right-hand side of \eqref{eq:ops_soc_cone_perp} is never larger than any of the terms in the right-hand side of \eqref{eq:ops_soc_ii_jj_ri}.

 We define a new optimization problem model with the improved voltage cone relaxation as:
 \begin{align}
    &\max &&  \mbox{Objective \eqref{eq:obj}} \nonumber\\
&\mbox{s.t.: \,\,\,}  \tag{SOC-OPS-P}
&& \mbox{Component relationships:}~\eqref{eq:ops_relationships}, \eqref{eq:z_bounds} \nonumber \\[-2pt]
&&& \mbox{Generation constraints:}~\eqref{eq:ops_active_gen_limits}, \eqref{eq:ops_reactive_gen_limits}  \nonumber\\[-2pt]
&&& \mbox{SOC power flow:}~\eqref{eq:ops_soc_active_power_balance}-\eqref{eq:ops_soc_angle_limits}, \eqref{eq:ops_soc_shunt_relaxation}, \eqref{eq:ops_soc_cone_perp}.
\nonumber
\end{align}

\section{Linearization of Conic Constraints} \label{sec:linearize_quads}
We aim to relax the SOC problem using linear constraints in order to capture some of the reactive power flow in the optimization problem to improve the solution quality over a DC-OPF formulation, but significantly improve the solution speed compared to the SOC-OPS problem.  To that end, we note that there are 3 nonlinear equations in the optimization problem:  \eqref{eq:ops_complex_thermal_limit_to}, \eqref{eq:ops_complex_thermal_limit_fr}, and \eqref{eq:ops_soc_cone_perp}.  We first look at the thermal constraints, then present two options for the voltage constraint.

\subsection{Linearize Thermal Limit:} The thermal limit  in eq. \eqref{eq:ops_complex_thermal_limit_fr}. has two quadratic terms $(P^L_{ij})^2$ and $(Q^L_{ij})^2$.  Let $y^P_{ij}$ and $y^Q_{ij}$ be two new variables that replace the quadratic terms in the thermal limit, 
\begin{equation}
    y^P_{ij} + y^Q_{ij} \le \boldsymbol{T}^2_{ij} z_{ij}  .\label{eq:linear_thermal_fr}
\end{equation}  

First we discuss the real power component.  We create a linear outer relaxation of the quadratic term by linearizing the quadratic at each point $l$ in a finite set of linearization points $L$:
\begin{equation}
    y^P_{ij} \ge 2 l P^L_{ij} - l^2 z_{ij} \quad \forall l \in L  . \label{eq:linear_fr_yp}  
\end{equation}

This is similarly done for the reactive power component:
\begin{equation}
    y^Q_{ij} \ge 2 l Q^L_{ij} - l^2 z_{ij} \quad \forall l \in L  . \label{eq:linear_fr_yq}
\end{equation}

We can use these constraints to replace the second order cone constraint of the thermal limit with several linear constraints.
The same approach is used to relax the eq. \eqref{eq:ops_complex_thermal_limit_to}, using the following constraints:
\begin{subequations}
    \begin{align}
            & y^P_{ji} + y^Q_{ji} \le \boldsymbol{T}_{ij} z_{ij} & \forall {ij} \in \mathcal{L} \\ 
            & y^P_{ji} \ge 2 l P^L_{ji} - l^2 z_{ij} & \forall l \in L, \quad \forall ij \in \mathcal{L}\\   
            & y^Q_{ji} \ge 2 l Q^L_{ji} - l^2 z_{ij} & \forall l \in L, \quad \forall ij \in \mathcal{L} .
    \end{align} \label{eq:linear_thermal_to}
\end{subequations}

We construct the following optimization model that uses the the linearized thermal constraints and the improved voltage cone from \ref{sec:tighten}
 \begin{align}
    &\max &&  \mbox{Objective \eqref{eq:obj}} \nonumber\\
&\mbox{s.t.: \,\,\,}  \tag{SOC-OPS-T}
&& \mbox{Component relationships:}~\eqref{eq:ops_relationships}, \eqref{eq:z_bounds} \nonumber \\[-2pt]
&&& \mbox{Generation constraints:}~\eqref{eq:ops_active_gen_limits}, \eqref{eq:ops_reactive_gen_limits}  \nonumber\\[-2pt]
&&& \mbox{SOC power flow:}~\eqref{eq:ops_soc_active_power_balance}, \eqref{eq:ops_soc_reactive_power_balance}, \eqref{eq:ops_soc_powerflow}-\eqref{eq:ops_soc_angle_limits}, \eqref{eq:ops_soc_shunt_relaxation}-\eqref{eq:linear_thermal_to}
\nonumber
\end{align}

\subsection{McCormick Voltage Relaxation} \label{sec:mccormick}
To create a fully linearized model we use the technique applied to the thermal limit and apply it to the voltage cone eq. \eqref{eq:ops_soc_cone_perp},   
$$\left(W^R_{ij}\right)^2 +  \left(W^I_{ij}\right)^2  \le W^{Fr}_{ij} W^{To}_{ij} .$$
We create new variables $y^{W^R}_{ij}$ and $y^{W^R}_{ij}$ to substitute for $W^R_{ij}$ and $W^I_{ij}$.  Then, we linearize the quadratic terms by adding cuts to
create the following new equations:
\begin{subequations}
    \begin{align}
        & y^{W^R}_{ij} \ge 2 l W^R_{ij} - l^2 z_{ij} & \forall l \in L \quad \forall ij \in \mathcal{L} \\
        & y^{W^I}_{ij} \ge 2 l W^I_{ij} - l^2 z_{ij} & \forall l \in L \quad \forall ij \in \mathcal{L} .
    \end{align} \label{eq:voltage_linear}
\end{subequations}
Adding these new variables to the voltage cone results in the following non-convex constraint:
\begin{equation}
    y^{W^R}_{ij}  +  y^{W^I}_{ij}  \le W^{Fr}_{ij} W^{To}_{ij} .\nonumber \label{eq:linear_lhs}
\end{equation}
We can relax this constraint by using a McCormick relaxation of the bi-linear product to create the following constraints:
\begin{subequations}
    \begin{align}
        & y^{W^R}_{ij}  \!\!\!\!\! +  y^{W^I}_{ij} \!\! \le W^{Fr}_{ij}  \overline{\boldsymbol{W^{To}}_{ij}} +  W^{To}_{ij} \underline{\boldsymbol{W^{Fr}}_{ij}} \!\!\! - \!\! \underline{\boldsymbol{W^{Fr}}_{ij}} \overline{\boldsymbol{W^{To}}_{ij}} z_{ij} \\    
        & y^{W^R}_{ij}  \!\!\!+  y^{W^I}_{ij} \!\! \le W^{Fr}_{ij} \underline{\boldsymbol{W^{To}}_{ij}} +  W^{To}_{ij} \overline{\boldsymbol{W^{Fr}}_{ij}} \!\!\!\!\! - \!\! \underline{\boldsymbol{W^{To}}_{ij}} \overline{\boldsymbol{W^{Fr}}_{ij}} z_{ij} .
    \end{align} \label{eq:voltage_mccormick}
\end{subequations}

A fully linear model is constructed using these equations:
 \begin{align}
    &\max &&  \mbox{Objective \eqref{eq:obj}} \nonumber\\
&\mbox{s.t.: \,\,\,}  \tag{SOC-OPS-M}
&& \mbox{Component relationships:}~\eqref{eq:ops_relationships}, \eqref{eq:z_bounds} \nonumber \\[-2pt]
&&& \mbox{Generation constraints:}~\eqref{eq:ops_active_gen_limits}, \eqref{eq:ops_reactive_gen_limits}  \nonumber\\[-2pt]
&&& \mbox{SOC power flow:}~\eqref{eq:ops_soc_active_power_balance}, \eqref{eq:ops_soc_reactive_power_balance}, \eqref{eq:ops_soc_powerflow}-\eqref{eq:ops_soc_angle_limits}, \eqref{eq:ops_soc_shunt_relaxation}, \eqref{eq:linear_thermal_fr}-\eqref{eq:voltage_mccormick}
\nonumber
\end{align}

\subsection{Secant Voltage Relaxation} \label{sec:secant}
We can create a linear relaxation of the of constraint \eqref{eq:ops_soc_cone_perp} that is tighter that the McCormick relaxation by using a secant bound.  

We first convert \eqref{eq:ops_soc_cone_perp} from the ``rotated second order cone'' into standard form:
\[ \Bigl[ \frac{1}{2} \bigl(W^{To}_{ij} + W^{Fr}_{ij}\bigr) \Bigr]^2 -
\Bigl[ \frac{1}{2} \bigl(W^{To}_{ij} - W^{Fr}_{ij}\bigr) \Bigr]^2 = W_{ij}^{Fr}W^{To}_{ij}. \]
Substituting into \eqref{eq:ops_soc_cone_perp} and rearranging yields the equivalent inequality:
\[ \left(W^R_{ij}\right)^2 +  \left(W^I_{ij}\right)^2 + \Bigl[ \frac{1}{2} \bigl(W^{To}_{ij} - W^{Fr}_{ij}\bigr) \Bigr]^2 \leq \Bigl[ \frac{1}{2} \bigl(W^{To}_{ij} + W^{Fr}_{ij}\bigr) \Bigr]^2. \]
We can use variables $y^{W^R}_{ij}$ and $y^{W^I}_{ij}$ to upper bound the terms $\left(W^R_{ij}\right)^2$ and  $\left(W^I_{ij}\right)^2$ with linear constraints as above. We can similarly introduce a variable $y^{WSum}_{ij}$ to upper bound the term
\[ \Bigl[ \frac{1}{2} \bigl(W^{To}_{ij} - W^{Fr}_{ij}\bigr) \Bigr]^2 \]
with linear constraints in the same way:
\begin{equation}
    y^{W^{Sum}}_{ij} \ge l \bigl(W^{To}_{ij} - W^{Fr}_{ij}\bigr) - l^2 z_{ij} \quad \forall l \in L \quad \forall ij \in \mathcal{L}. \label{eq:y_w_sum}
\end{equation}
What remains is a quadratic term on the right-hand side. We can use a secant upper bound for that.  To simply the derivation,  we first define
\begin{equation}
    s_{ij} = \frac{1}{2} \bigl(W^{To}_{ij} + W^{Fr}_{ij}\bigr) \nonumber \label{eq:secant_s}
\end{equation}
and define its lower and upper bounds as
\begin{align*}
\boldsymbol{\ell_{ij}} &= \frac{1}{2} (\boldsymbol{\underline{V}}_i^2 + \boldsymbol{\underline{V}}_j^2) \\
\boldsymbol{u_{ij}} &= \frac{1}{2} (\boldsymbol{\bar{V}}_i^2 + \boldsymbol{\bar{V}}_j^2)
\end{align*}
Then the secant upper bound is defined by the line connecting the points $(\boldsymbol{\ell}_{ij},\boldsymbol{\ell}_{ij}^2)$ and $(\boldsymbol{u}_{ij},\boldsymbol{u}_{ij}^2)$. 
The line has slope $(\ell_{ij} + u_{ij})$ and $y$-intercept $-\ell_{ij} u_{ij}$. 
Finally, when using the secant upper bound, the $y$-intercept can be multiplied with $z_{ij}$ since all terms in the inequality go to zero when $z_{ij} = 0$. 
Substituting this upper bound on the quadratic term and using the definition of $s_{ij}$ this leads to the linear inequality:
\begin{equation}
    y_{ij}^{W^R} \!\!\!\!\! + y_{ij}^{W^I} \!\!\!\! + y_{ij}^{W^{Sum}} \!\! \leq  (\boldsymbol{u}_{ij} \! + \boldsymbol{\ell}_{ij}) \frac{W_{ij}^{Fr} \!\! + W_{ij}^{To}}{2} -\boldsymbol{\ell}_{ij} \boldsymbol{u}_{ij} z_{ij} \label{eq:secant_bound}
\end{equation}

The linear model using a secant bound for the SOC voltage cone is:
 \begin{align}
    &\max &&  \mbox{Objective \eqref{eq:obj}} \nonumber\\
&\mbox{s.t.: \,\,\,}  \tag{SOC-OPS-S}
&& \mbox{Component relationships:}~\eqref{eq:ops_relationships}, \eqref{eq:z_bounds} \nonumber \\[-2pt]
&&& \mbox{Generation constraints:}~\eqref{eq:ops_active_gen_limits}, \eqref{eq:ops_reactive_gen_limits}  \nonumber\\[-2pt]
&&& \mbox{SOC power flow:}~\eqref{eq:ops_soc_active_power_balance}, \eqref{eq:ops_soc_reactive_power_balance}, \eqref{eq:ops_soc_powerflow}-\eqref{eq:ops_soc_angle_limits}, \eqref{eq:ops_soc_shunt_relaxation}, \nonumber \\ &&&\eqref{eq:linear_thermal_fr}-\eqref{eq:voltage_linear},\eqref{eq:y_w_sum}, \eqref{eq:secant_bound}
\nonumber
\end{align}

\section{Case Study} \label{sec:case_study}
We solve the OPS problem with each of the formulations presented in this work, in addition to the DC power flow model in \cite{haag2024long}. We solve each formulation on cases from PGLib \cite{pglib} ranging from  5 buses to 118 buses.  Each problem is solved with Gurobi v11 \cite{gurobi}, using the PowerModelsWildfire.jl framework \cite{rhodes2020balancing}, in the Julia programming language \cite{julia}, on a system with 2 64 core AMD EPYC 7H12 processor.  For each network, we solve 5 randomly generated wildfire risk scenarios.  To evaluate solution quality, we fix the binary variables for each solution and re-solve the generators set points to maximize load delivery with the exact SOC power flow, known as the SOC-Redispatch problem from \cite{haag2024long}.

Note that the 89 bus test case has some negative loads. The problem objective in maximizing load is not well-behaved with these loads. We set these loads to 0, but alternatives such as fixing these load values or changing the negative loads to generators are also viable options.

\section{Results} \label{sec:results}
We explore three key results from this work.  First, we examine the computational speed improvements of the presented formulations in comparison to the SOC and DC power flow formulations in \cite{rhodes2020balancing}.  Next we study the solution quality of the decisions made by each formulation.  Finally, we examine the impact of increasing the number of linear cuts that represent the $P^2$, $Q^2$, ${(W^R)}^2$, ${(W^I)}^2$, and ${(W^{Sum})}^2$ variables.

\subsection{Computational Speed}

\begin{figure}[t]
    \centering
    \includegraphics[width=\columnwidth]{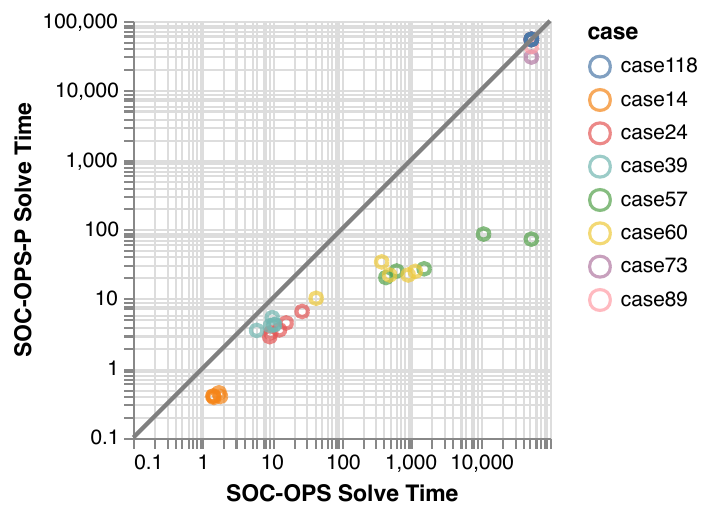}
    \caption{\small Comparison of SOC-OPS and SOC-OPS-P solve time. The SOC-OPS-P problem is faster than the SOC-OPS problem, especially on the 57 and 60 bus cases. The 73, 89, 118 cases generally terminate at the time limit.}
    \label{fig:solve_time}
\end{figure}

\begin{table*}[t]
\centering
    \caption{\small Average solution time [sec].  Parentheses indicates number of scenarios that terminate at the time limit.}
    \label{tab:solve_speed}
\begin{tabular}{@{}rrcrcrcrcrcrc@{}}
\cmidrule[\heavyrulewidth](l){1-13}
\multicolumn{1}{c}{} &
  \multicolumn{2}{c}{SOC-OPS} &
  \multicolumn{2}{c}{SOC-OPS-P} &
  \multicolumn{2}{c}{SOC-OPS-T} &
  \multicolumn{2}{c}{SOC-OPS-M} &
  \multicolumn{2}{c}{SOC-OPS-S} &
  \multicolumn{2}{c}{DC-OPS} \\ 
  \cmidrule(lr){2-3} \cmidrule(lr){4-5} \cmidrule(lr){6-7} \cmidrule(lr){8-9}  \cmidrule(lr){10-11}  \cmidrule(lr){12-13} 
\multicolumn{1}{c}{Case} &
  \multicolumn{1}{l}{\begin{tabular}[c]{@{}l@{}}Avg Solve \\ Time (s)\end{tabular}} &
  \multicolumn{1}{r}{\begin{tabular}[c]{@{}r@{}}Time \\ Limit\end{tabular}} &
  \multicolumn{1}{l}{\begin{tabular}[c]{@{}l@{}}Avg Solve \\ Time (s)\end{tabular}} &
  \multicolumn{1}{r}{\begin{tabular}[c]{@{}r@{}}Time \\ Limit\end{tabular}} &
  \multicolumn{1}{l}{\begin{tabular}[c]{@{}l@{}}Avg Solve \\ Time (s)\end{tabular}} &
  \multicolumn{1}{r}{\begin{tabular}[c]{@{}r@{}}Time \\ Limit\end{tabular}} &
  \multicolumn{1}{l}{\begin{tabular}[c]{@{}l@{}}Avg Solve \\ Time (s)\end{tabular}} &
  \multicolumn{1}{r}{\begin{tabular}[c]{@{}r@{}}Time \\ Limit\end{tabular}} &
  \multicolumn{1}{l}{\begin{tabular}[c]{@{}l@{}}Avg Solve \\ Time (s)\end{tabular}} &
  \multicolumn{1}{r}{\begin{tabular}[c]{@{}r@{}}Time \\ Limit\end{tabular}} &
  \multicolumn{1}{l}{\begin{tabular}[c]{@{}l@{}}Avg Solve \\ Time (s)\end{tabular}} &
  \multicolumn{1}{r}{\begin{tabular}[c]{@{}r@{}}Time \\ Limit\end{tabular}} \\ 
   \cmidrule(lr){1-1} \cmidrule(lr){2-3} \cmidrule(lr){4-5} \cmidrule(lr){6-7} \cmidrule(lr){8-9}  \cmidrule(lr){10-11}  \cmidrule(lr){12-13}
case14 &
  1.56 &
  0 &
  0.40 &
  0 &
  0.58 &
  0 &
  0.46 &
  0 &
  0.51 &
  0 &
  0.08 &
  0 \\
case24 &
  14.87 &
  0 &
  4.12 &
  0 &
  6.00 &
  0 &
  1.77 &
  0 &
  2.64 &
  0 &
  0.43 &
  0 \\
case39 &
  9.36 &
  0 &
  4.27 &
  0 &
  5.89 &
  0 &
  2.44 &
  0 &
  5.15 &
  0 &
  0.34 &
  0 \\
case57 &
  13530.08 &
  1 &
  45.75 &
  0 &
  57.57 &
  0 &
  33.33 &
  0 &
  78.09 &
  0 &
  1.06 &
  0 \\
case60 &
  593.20 &
  0 &
  22.46 &
  0 &
  23.94 &
  0 &
  10.79 &
  0 &
  7.62 &
  0 &
  1.05 &
  0 \\
case73 &
  54001.53 &
  5 &
  49209.94 &
  4 &
  45053.47 &
  4 &
  2445.67 &
  0 &
  20547.61 &
  1 &
  90.39 &
  0 \\
case89 &
  54000.74 &
  5 &
  51702.86 &
  4 &
  54000.08 &
  5 &
  499.72 &
  0 &
  1107.83 &
  0 &
  152.28 &
  0 \\
case118 &
  54000.96 &
  5 &
  54001.05 &
  5 &
  54000.71 &
  5 &
  50531.38 &
  4 &
  54000.05 &
  5 &
  182.99 &
  0 \\ \bottomrule
\end{tabular}
\end{table*}

Table \ref{tab:solve_speed} shows the average solution time of each problem across five scenarios for each power grid case.  Numbers in parentheses show how many of the five scenarios terminated at the time-limit, rather than optimality. 
The SOC-OPS-P reformulation shows a distinct speedup compared to the SOC-OPS problem on systems of 60 buses or smaller, but on larger systems it still terminates at the time limit. The SOC-OPS-T, which replaces 2 SOC constraints with linear cuts, does not show any improvement in solution speed over the SOC-OPS-P formulation.   

The fully linear models SOC-OPS-M and SOC-OPS-S are significantly faster than the SOC-OPS-P problem, but struggle to converge within 15 hours when the system size is 118 buses.  Meanwhile the DC-OPS problem can solve the 118 bus system in 183 seconds.  

Figure \ref{fig:solve_time} shows a more detailed runtime comparison of the SOC-OPS and the SOC-OPS-P formulations, with SOC-OPS solvetimes on the x-axis and SOC-OPS-P solvetimes on the y-axis. Each point represent one test case with a specific wildfire risk scenario, and points of similar color refer to solutions for the same test case. If a point is on the black line, it indicates that the SOC-OPS and SOC-OPS-P have the same solve time. Each point below the line shows the SOC-OPS-P problem is faster for that scenario, and the runtime improvement is especially strong on the 57 bus and 60 bus grid models.  The runtime of the SOC-OPS-P problem does not vary much across the 5 scenarios of each network, while the SOC-OPS runtime varies by almost  two orders of magnitude depending on the risk scenario.

\subsection{SOC Accuracy}
Next we examine if the alternative formulations provide sufficiently accurate solutions, compared to the exact SOC-OPS formulation.

In Table \ref{tab:soc_bound}, we show the ratio of the solution objective value of each method over the best solution bound found by either the SOC-OPS or SOC-OPS-P formulations.  As should be expected, the value is 100\% for SOC-OPS and SOC-OPS-P when the solution terminates as optimal, but is less than 100\% when the solution terminates early.  Near the time limit, the solutions are still high-quality, but the optimality gap has not fully closed.  The solutions to the relaxations are all greater than 100\%, indicating that those formulations find overly optimistic solutions.  The SOC-OPS-M has the largest relaxation, and SOC-OPS-S improves over this.  The SOC-OPS-T relaxation is surprisingly tight, and hardly exceeds the best bound from the SOC solutions.  
The objective value of the DC-OPS approximation solutions are  better than the the SOC-OPS-S formulation, except in the 14, 24, and 73 bus cases. However, the solutions are still very close and it should be recalled that SOC-OPS-S provably provides a relaxation, whereas DC-OPS is an approximation without such a guarantee.

\begin{table*}[t]
    \centering
    \caption{\small Ratio of solution objective value over the best SOC bound.  Solutions over 100\% represent an overestimate of the load delivered and wildfire risk reduced.}
    \label{tab:soc_bound}
    \input{tables/soc_bound}
\end{table*}

Next we focus on the load delivery component of the objective function which is most impacted by the choice of power flow formulation.  Table \ref{tab:redispatch_performance}, which shows the ratio of load delivered by the SOC Redispatch solution over the estimated load delivered in the SOC OPS problem variant, averaged across all scenarios that have feasible SOC Redispatch power flow solution.  The number in parentheses shows the number of scenarios included in the average.
Here, the 14-bus case is the worst-case power grid for any of the relaxations, with typical redispatch performance in the range of 50\% to 70\% for any of the relaxations.  SOC-OPS-S shows significantly better performance than the other linear formulations SOC-OPS-M and DC-OPS for this worst case network.

Across all scenarios, the SOC-OPS-T problem finds solutions with near identical load delivery as the SOC solutions.
Of the two linear relaxations, the McCormick relaxation performs worse on all networks than the Secant relaxation, a trade-off for the faster runtime shown in Table \ref{tab:solve_speed}.  The Secant relaxation has worse performance than the DC-OPS problem in five of the networks.

\begin{table*}[t]
\centering
\caption{\small Average Redispatch Performance of Feasible Solutions. Values below 100\% indicate an overestimate of the load delivered. Parentheses indicate the number of scenarios included in the average.} \label{tab:redispatch_performance}
\input{tables/soc_quality}
\end{table*}

\subsection{Effect of increasing linearization points}
\begin{figure}[t]
    \centering
    \includegraphics[width=\columnwidth]{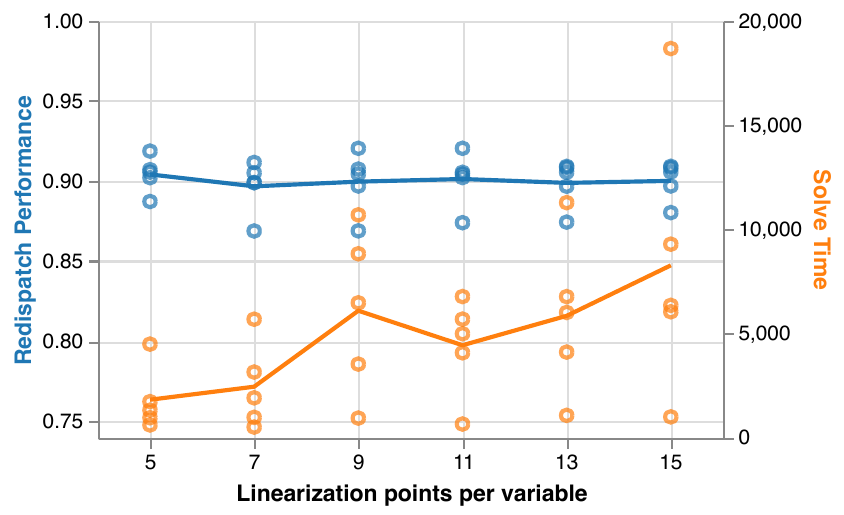}
    \caption{\small Redispatch Performance and Solve Time of SOC-OPS-M on the 89 bus system as a the number of linearization points are varied. Points represent individual scenarios, and the line is the average value.} \label{fig:l_tradeoff}
\end{figure}

Last, we study the impact of adjusting the number of linear cuts on the trade-off of solution speed and solution quality in a case study involving the 73 bus system.  
Figure  \ref{fig:l_tradeoff} shows the change in SOC Redispatch load delivery and and the solve time of the SOC-OPS-M problem as the number of linear cuts per quadratic relaxation increases from 5 to 15.

As the number of cuts is increased from 5 to 15, the solution time increased by around 2.5x.  At the same time, no significant change in redispatch performance occurs, with the performance staying around 0.9 regardless of the number of cuts.  

Increasing the the number of cuts does not have an impact on the tightness of this relaxation, indicating that the low quality solutions are not related to the voltage values exceeding the quadratic terms.  Rather, the likely cause is the McCormick relaxation of $W^{Fr}_{ij}W^{To}_{ij}$ is too loose, allowing solutions that do not recover to good SOC-feasible solutions.

\section{Discussion}
\label{sec:discussion}
The goal of this work was to  relax the SOC-OPS problem to a linear problem with fast performance, but reasonable accuracy of the ability to deliver load.  Unfortunately, this goal was not achieved, but there are still some interesting results.

First, the SOC-OPS-P reformulation represents a significant speedup over the original SOC-OPS problem formulation. This reformulation can be applied to any topology-switching SOC power flow formulation problem.

Second, we expected the SOC-OPS-T problem would improve solution speed by reducing the number of SOC constraints from $3|\mathcal{B}|$ to $|\mathcal{B}|$. In our numerical case study, there was no speedup from this change, but also there is hardly an impact on solution quality indicating that this relaxation may have some value if an effective method of relaxing the SOC voltage cone can be found.

Third, the Secant model provided a tighter relaxation of the SOC voltage cone than the McCormick model, and translated to improved redispatch quality. Among the linear models, the Secant model found the best quality solution, with the best redispatch performance, on the worst case network, the 14 bus network. 
While the solve time of the SOC-OPS-S formulation is dominated by the DC-OPS model, the SOC-OPS-S formulation is a relaxation and provides a bound on the solution quality.  

Fourth, increasing the number of linearization points had no impact on solution quality.  This indicates that the poor quality is to due the relaxation of the right-hand side of the voltage cone in eq. \eqref{eq:ops_soc_cone_perp}.  Improving the tightness, for example with a piecewise linear relaxation, may lead to improved quality, but the speed will suffer as more binary variables are added to the problem.

Fifth, the speed of the DC-OPS problem is remarkably fast.  The increased number of (linear) constraints in the SOC-OPS problem, in addition to constraints added by the linearization of the quadratic terms, limits an SOC-OPS relaxation to systems smaller than 118 buses.

\section{Conclusion} \label{sec:conclusion}
In this work, we explore the SOC-OPS problem and introduce several reformulations and relaxations of the problem to identify a reasonable trade-off between maintaining good solution quality and increasing solve time.  We show that a simple reformulation of the problem can decrease solve time from 13,500 seconds to 45 seconds.  However, additional efforts to relax the problem to linear formulations were not successful at finding a good trade-off for reduced solution time.  The solution quality of those linearized solutions are no better than the DC-OPS problem, but with significantly increased solve time. 

Future work may include more investigation on the right hand side of eq. \eqref{eq:ops_soc_cone_perp}, where the relaxation is weakest.  Improving the relaxation with a piecewise linear model may help, or exploring alternative modeling formulations.  
Additionally, work could start from the DC-OPS problem and focus on reducing the very high inaccuracy of the power flow formulation in these large scale outage scenarios.

\vspace{+5pt}
\small \noindent Los Alamos Unlimited Release LA-UR-25-20527. Reviewed for release outside the Laboratory with no distribution restrictions.

\bibliographystyle{IEEEtran}
\bibliography{IEEEabrv,references}

\end{document}

%% file: tables/soc_bound.tex
\begin{tabular}{@{}rrrrrrr@{}}
\toprule
Case & SOC-OPS & SOC-OPS-P & SOC-OPS-T & SOC-OPS-M & SOC-OPS-S & DC-OPS\\ \midrule
case14 & 100.00\% & 100.00\% & 100.00\% & 103.01\% & 102.49\% & 102.84\%\\
case24 & 100.00\% & 100.00\% & 100.00\% & 101.80\% & 100.88\% & 100.93\%\\
case39 & 100.00\% & 100.00\% & 100.01\% & 101.47\% & 100.54\% & 100.25\%\\
case57 & 100.00\% & 100.00\% & 100.00\% & 101.36\% & 100.78\% & 100.70\%\\
case60 & 100.00\% & 100.00\% & 100.02\% & 101.05\% & 100.60\% & 100.22\%\\
case73 & 99.70\% & 99.80\% & 99.82\% & 101.64\% & 100.63\% & 100.91\%\\
case89 & 98.35\% & 99.81\% & 99.78\% & 102.33\% & 101.90\% & 101.67\%\\
case118 & 98.65\% & 99.00\% & 99.08\% & 100.90\% & 100.39\% & 100.13\%\\
\bottomrule
\end{tabular}

%% file: tables/soc_quality.tex
\begin{tabular}{@{}rrrrrrr@{}}
\toprule
Case & SOC-OPS & SOC-OPS-P & SOC-OPS-T & SOC-OPS-M & SOC-OPS-S & DC-OPS\\ \midrule
case14 & 100.00\% (5) & 100.00\% (5) & 100.00\% (5) & 49.39\% (5) & 68.39\% (5) & 49.49\% (5)\\
case24 & 100.00\% (5) & 100.00\% (5) & 100.00\% (5) & 91.48\% (5) & 96.35\% (5) & 94.28\% (5)\\
case39 & 100.11\% (5) & 100.11\% (5) & 100.11\% (5) & 89.35\% (5) & 96.14\% (5) & 99.37\% (5)\\
case57 & 100.00\% (5) & 100.00\% (5) & 100.00\% (5) & 67.51\% (1) & 85.57\% (5) & 89.52\% (5)\\
case60 & 100.00\% (5) & 100.00\% (5) & 99.96\% (5) & 92.84\% (5) & 97.95\% (5) & 98.91\% (5)\\
case73 & 100.00\% (5) & 100.00\% (3) & 99.98\% (5) & 89.67\% (5) & 96.49\% (5) & 93.23\% (5)\\
case89 & 100.00\% (2) & 100.00\% (4) & 100.00\% (4) & 71.93\% (5) & 80.27\% (5) & 84.90\% (5)\\
case118 & 100.00\% (5) & 100.00\% (3) & 99.99\% (5) & 72.64\% (2) & 81.44\% (5) & 82.66\% (5)\\
\bottomrule
\end{tabular}

%% file: ieee_conf.bbl
\begin{thebibliography}{10}
\providecommand{\url}[1]{#1}
\csname url@samestyle\endcsname
\providecommand{\newblock}{\relax}
\providecommand{\bibinfo}[2]{#2}
\providecommand{\BIBentrySTDinterwordspacing}{\spaceskip=0pt\relax}
\providecommand{\BIBentryALTinterwordstretchfactor}{4}
\providecommand{\BIBentryALTinterwordspacing}{\spaceskip=\fontdimen2\font plus
\BIBentryALTinterwordstretchfactor\fontdimen3\font minus \fontdimen4\font\relax}
\providecommand{\BIBforeignlanguage}[2]{{%
\expandafter\ifx\csname l@#1\endcsname\relax
\typeout{** WARNING: IEEEtran.bst: No hyphenation pattern has been}%
\typeout{** loaded for the language `#1'. Using the pattern for}%
\typeout{** the default language instead.}%
\else
\language=\csname l@#1\endcsname
\fi
#2}}
\providecommand{\BIBdecl}{\relax}
\BIBdecl

\bibitem{castillo2014risk}
A.~Castillo, ``Risk analysis and management in power outage and restoration: A literature survey,'' \emph{Electric Power Systems Research}, vol. 107, pp. 9--15, 2014.

\bibitem{qiu2017integrated}
F.~Qiu and P.~Li, ``An integrated approach for power system restoration planning,'' \emph{Proceedings of the IEEE}, vol. 105, no.~7, pp. 1234--1252, 2017.

\bibitem{rhodes2020balancing}
N.~Rhodes, L.~Ntaimo, and L.~Roald, ``Balancing wildfire risk and power outages through optimized power shut-offs,'' \emph{IEEE Trans. on Power Syst.}, vol.~36, no.~4, pp. 3118--3128, 2020.

\bibitem{mld}
C.~Coffrin, R.~Bent, B.~Tasseff, K.~Sundar, and S.~Backhaus, ``Relaxations of ac maximal load delivery for severe contingency analysis,'' \emph{IEEE Trans. on Power Syst.}, vol.~34, no.~2, pp. 1450--1458, March 2019.

\bibitem{van2015transmission}
P.~Van~Hentenryck and C.~Coffrin, ``Transmission system repair and restoration,'' \emph{Mathematical Programming}, vol. 151, pp. 347--373, 2015.

\bibitem{haag2024long}
E.~Haag, N.~Rhodes, and L.~Roald, ``Long solution times or low solution quality: On trade-offs in choosing a power flow formulation for the optimal power shutoff problem,'' \emph{Electric Power Systems Research}, vol. 234, p. 110713, 2024.

\bibitem{rhodes2021powermodelsrestoration}
N.~Rhodes, D.~M. Fobes, C.~Coffrin, and L.~Roald, ``Powermodelsrestoration. jl: An open-source framework for exploring power network restoration algorithms,'' in \emph{2021 Power Syst. Comput. Conf.}, 2021.

\bibitem{kody2022sharing}
A.~Kody, A.~West, and D.~K. Molzahn, ``Sharing the load: Considering fairness in de-energization scheduling to mitigate wildfire ignition risk using rolling optimization,'' in \emph{2022 IEEE 61st Conference on Decision and Control (CDC)}.\hskip 1em plus 0.5em minus 0.4em\relax IEEE, 2022, pp. 5705--5712.

\bibitem{rhodes2024security}
N.~Rhodes, C.~Coffrin, and L.~Roald, ``Security constrained optimal power shutoff for wildfire risk mitigation,'' \emph{IET Generation, Transmission \& Distribution}, vol.~18, no.~18, pp. 2972--2986, 2024.

\bibitem{rhodes2023co}
N.~Rhodes and L.~A. Roald, ``Co-optimization of power line shutoff and restoration under high wildfire ignition risk,'' in \emph{2023 IEEE Belgrade PowerTech}.\hskip 1em plus 0.5em minus 0.4em\relax IEEE, 2023, pp. 1--7.

\bibitem{taylor2022framework}
S.~Taylor and L.~A. Roald, ``A framework for risk assessment and optimal line upgrade selection to mitigate wildfire risk,'' \emph{2022 Power Syst. Comput. Conf.}, 2022.

\bibitem{bertoletti2022transmission}
A.~Z. Bertoletti and J.~C. do~Prado, ``Transmission system expansion planning under wildfire risk,'' in \emph{2022 North Am. Power Symp.}, 2022.

\bibitem{bayani2023resilient}
R.~Bayani and S.~D. Manshadi, ``Resilient expansion planning of electricity grid under prolonged wildfire risk,'' \emph{IEEE Trans. on Smart Grid}, 2023.

\bibitem{kody2022optimizing}
A.~Kody, R.~Piansky, and D.~K. Molzahn, ``Optimizing transmission infrastructure investments to support line de-energization for mitigating wildfire ignition risk,'' \emph{arXiv preprint arXiv:2203.10176}, 2022.

\bibitem{astudillo2022managing}
A.~Astudillo, B.~Cui, and A.~S. Zamzam, ``Managing power systems-induced wildfire risks using optimal scheduled shutoffs,'' National Renewable Energy Lab, Golden, CO (United States), Tech. Rep., 2022.

\bibitem{yang2022resilient}
W.~Yang, S.~N. Sparrow, M.~Ashtine, D.~C. Wallom, and T.~Morstyn, ``Resilient by design: Preventing wildfires and blackouts with microgrids,'' \emph{Applied Energy}, vol. 313, p. 118793, 2022.

\bibitem{hanna2021optimal}
R.~Hanna, ``Optimal investment in microgrids to mitigate power outages from public safety power shutoffs,'' in \emph{2021 IEEE Power \& Energy Society General Meeting (PESGM)}.\hskip 1em plus 0.5em minus 0.4em\relax IEEE, 2021, pp. 1--5.

\bibitem{yang2024multistage}
H.~Yang, H.~Yang, N.~Rhodes, L.~Roald, and L.~Ntaimo, ``Multistage stochastic program for mitigating power system risks under wildfire disruptions,'' \emph{Power Systems Computational Conference}, 2024.

\bibitem{yang2024multi}
H.~Yang, N.~Rhodes, H.~Yang, L.~Roald, and L.~Ntaimo, ``Multi-period power system risk minimization under wildfire disruptions,'' \emph{IEEE Transactions on Power Systems}, 2024.

\bibitem{zhou2024mitigating}
Y.~Zhou, K.~Sundar, H.~Zhu, and D.~Deka, ``Mitigating the impact of uncertain wildfire risk on power grids through topology control,'' in \emph{2024 18th International Conference on Probabilistic Methods Applied to Power Systems (PMAPS)}.\hskip 1em plus 0.5em minus 0.4em\relax IEEE, 2024, pp. 1--6.

\bibitem{hong2022data}
W.~Hong, B.~Wang, M.~Yao, D.~Callaway, L.~Dale, and C.~Huang, ``Data-driven power system optimal decision making strategy underwildfire events,'' Lawrence Livermore National Lab, Tech. Rep., 2022.

\bibitem{pglib}
{PGLib Optimal Power Flow Benchmarks}, ``The ieee pes task force on benchmarks for validation of emerging power system algorithms,'' Published online at \url{https://github.com/power-grid-lib/pglib-opf}, 2021.

\bibitem{gurobi}
{Gurobi Optimization, Inc.}, ``Gurobi optimizer reference manual,'' Published online at \url{http://www.gurobi.com}, 2014.

\bibitem{julia}
\BIBentryALTinterwordspacing
J.~Bezanson, A.~Edelman, S.~Karpinski, and V.~Shah, ``Julia: A fresh approach to numerical computing,'' \emph{SIAM Rev.}, vol.~59, no.~1, pp. 65--98, 2017. [Online]. Available: \url{https://doi.org/10.1137/141000671}
\BIBentrySTDinterwordspacing

\end{thebibliography}
